\documentclass[fleqn,prl,reprint,aps]{revtex4-1}
\usepackage{float,graphicx,amsmath,amssymb,booktabs,subfigure,hyperref,pifont,color}

\newcommand{\del}{\partial}

\renewcommand{\v}[1]{\mathbf{#1}}

\newcommand{\ave}[1]{\langle {#1} \rangle}

\begin{document}
\title{Still water: dead zones and collimated ejecta from the impact of granular jets}
\author{Jake Ellowitz, Herv\'{e} Turlier, Nicholas Guttenberg, Wendy W.\ Zhang, Sidney R.\ Nagel}
\affiliation{The James Franck Institute and the Department of Physics, University of Chicago, Chicago, IL 60637}
\date{\today}

\begin{abstract}
When a dense granular jet hits a target, it forms a large dead zone and ejects a highly collimated conical sheet with a well-defined opening angle. Using experiments, simulations, and continuum modeling, we find that this opening angle is insensitive to the precise target shape and the dissipation mechanisms in the flow. We show that this surprising insensitivity arises because dense granular jet impact, though highly dissipative, is nonetheless controlled by the limit of perfect fluid flow.
\end{abstract}

\maketitle

Students are familiar with liquids as an intermediate state of matter: like gases, they flow easily but, like solids, they are condensed due to inter-particle attractions. Later they may be taught that liquids can be modeled without attractions if the particle density is kept high by confinement~\cite{van_der_waals,rigid_sphere,kudrolli}. However, even without attractions or confinement, non-cohesive particles can behave like a liquid: when a high-density jet of grains hits a target it ejects particles in a thin sheet similar to that created by the impact of a liquid jet~\cite{cheng,clanet}. Here, we investigate why these two different types of materials behave in such a similar way.

In our experiments we measured the velocity near the impact center of the dense granular jet. Our measurements reveal a dead zone, a region of nearly immobile particles, instead of the smoothly varying straining flow characteristic of Newtonian liquid impact. We then used discrete-particle simulations to examine how this qualitative change in the velocity field alters the ejecta and found that it is only weakly modified. To understand the origin of this insensitivity we first used experiment and simulation to show that the granular motion in this regime can be modeled by an incompressible frictional flow, then analyzed the behavior of the ejecta in the continuum model in the limit where the dissipation vanishes. We find that the form of the ejecta is dictated by inertia and therefore highly robust. The angle of the emerging ejecta is insensitive to the type of dissipation mechanisms present, whether the central region is flowing or static, and the target shape. 

This study shares with experiments at the Relativistic Heavy Ion Collider (RHIC) an interest relating the ejecta pattern to bulk properties inside the impact zone~\cite{rhic_nature}.  Those experiments show that the collisions of gold ions produce surprisingly collimated scattering patterns. Some researchers have interpreted this coherent ejecta as evidence that the quark gluon plasma forms a dense, nearly perfect liquid~\cite{rhic_euler,song_viscosity,jacak_rhic}. Granular jets that also scatter like liquids constitute a macroscopic analog. In granular physics, low-speed, dense flows~\cite{jaeger_sandpile,lube_huppert_sandpile} and high-speed, dilute flows~\cite{sand_dune,gray_shock,granular_stream,swinney_shock} have received much attention, but less is known about the high-speed, dense regime we examine here.  This regime is relevant in a wide variety of contexts.  Two examples are protoplanetary formation by the collision of dust aggregates~\cite{blum_planetary,johansen_planetary,teiser_planetary,just_blum,langkowski_teiser_blum}, and abrasive blasting using high speed sand jets~\cite{sandblast1,sandblast2}.

\begin{figure}
\centering
\includegraphics[width=\columnwidth]{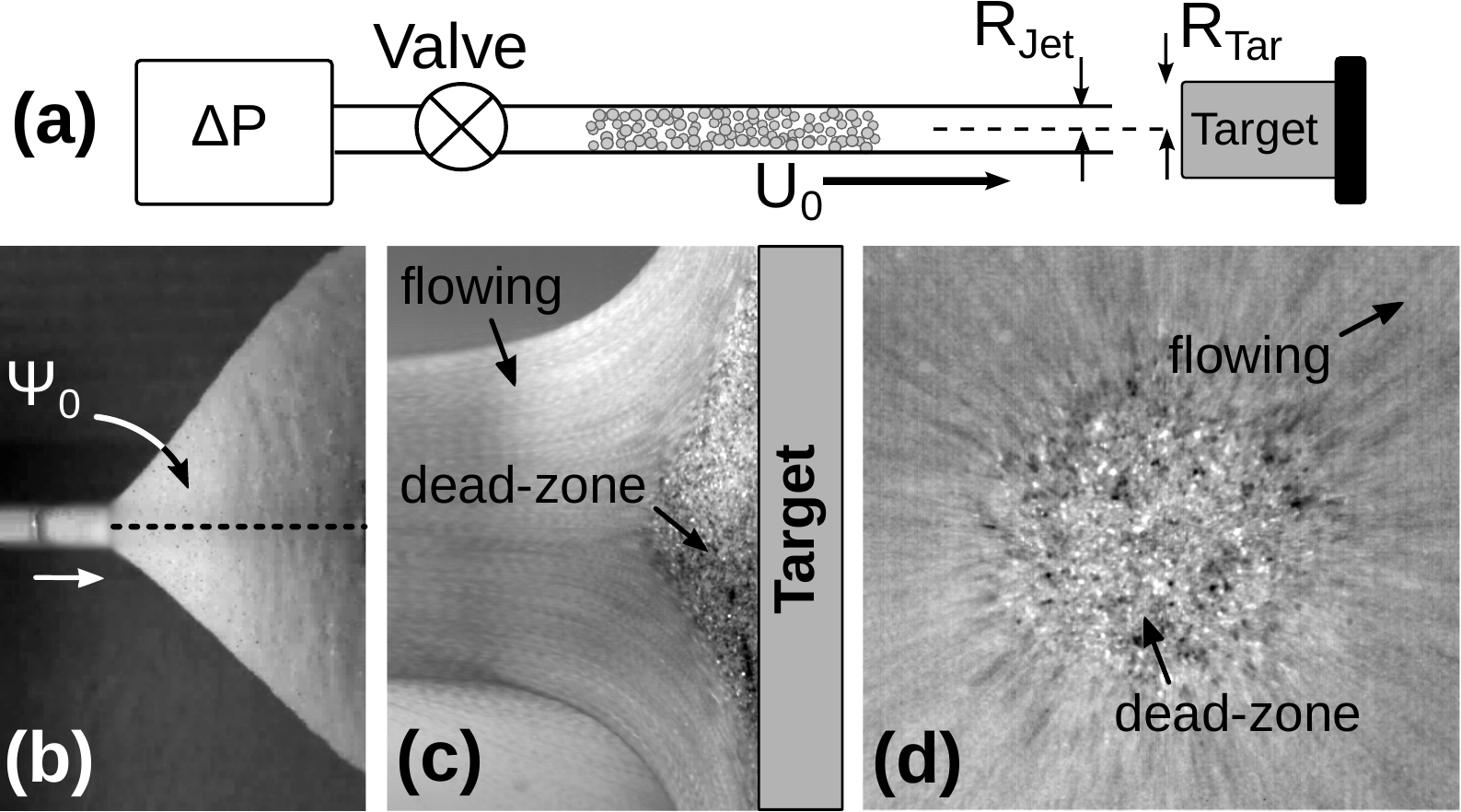}
\caption{\label{fig:exp}Experiments reveal that a large dead-zone forms during granular jet impact. {\bf (a)} Schematic: a burst of air ejects a dense column of spherical glass beads from a long tube. The granular jet subsequently hits a target at a speed $U_0$. {\bf (b)} Impact against the target produces an axisymmetric ejecta sheet with opening angle $\Psi_0$. 
Long-time exposure of granular jet impact as viewed from {\bf (c)}, the side, and {\bf (d)}, below, reveals a large dead-zone.
}
\end{figure}

In our experiment, depicted in Fig.~\ref{fig:exp}{\bf (a)}, we follow the protocol of Cheng {\em et al}~\cite{cheng}. High-pressure gas pushes a dense plug of non-cohesive glass beads of radius $R_{\rm G} = 54\pm 9\,\mu{\rm m}$ out of a tube of radius $R_{\rm Jet} = 0.8\,{\rm cm}$ at a speed $U_0$. Depending on the gas pressure, $U_0$ is between $1\,{\rm m/s}$ and $16\,{\rm m/s}$. This jet of particles hits a target of radius $R_{\rm Tar}$. 
 The particles are then ejected from the target at an angle $\Psi_0$ in a thin axisymmetric cone (Fig.~\ref{fig:exp}{\bf (b)}). 

\begin{figure}
\centering
\includegraphics[width=0.75\columnwidth]{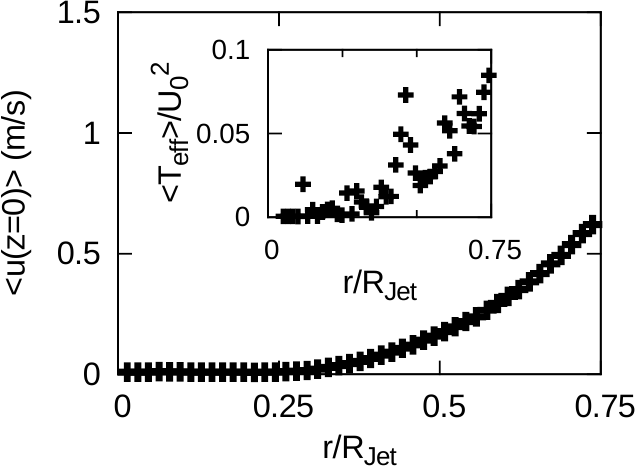}
\caption{\label{fig:expdz}
PIV measurements of the azimuthally averaged radial speed of grains at the transparent target $\langle u_r(r,z=0)\rangle$ measured from below.  
We see a dead zone at the target center. Inset shows the effective granular temperature $T_{\rm eff}$, or the velocity fluctuations, normalized by $U_0^2$. The dead zone is not only static but also cold. 
}
\end{figure}

We present measurements of the internal flow for impact onto a roughened steel target with radius $R_{\rm Tar} = R_{\rm Jet}$ and onto a smooth, transparent glass target with radius $R_{\rm Tar} = 6.4 R_{\rm Jet}$.
We view the impact zone near the steel target by slicing the originally axisymmetric experiment in half along its length and observing the central region from the side through a glass window (Fig.~\ref{fig:exp}{\bf (c)}), and view the impact zone near the transparent target by observing from the rear (Fig.~\ref{fig:exp}{\bf (d)}). 
The interior of the jet, as viewed from both the side and rear, reveal a region extending over a significant fraction of the target where the grain motion is negligible compared to the surrounding flow. We call this the dead zone. 

\begin{figure}
\centering
\includegraphics[width=0.8\columnwidth]{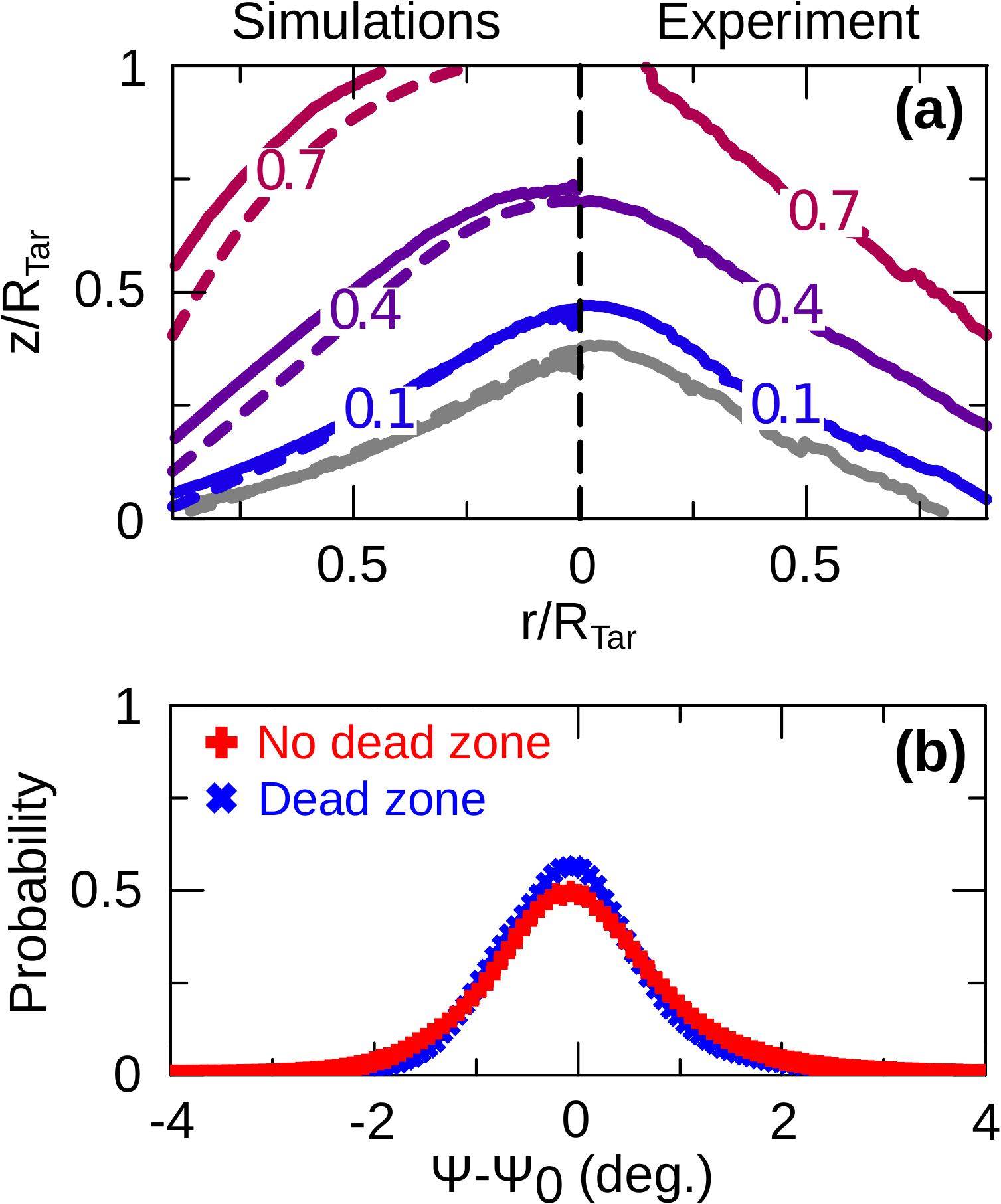}
\caption{\label{fig:dem}
(color online) Discrete-particle simulations using a frictional target reproduce the velocity field from experiments with roughened targets. Using a frictionless target in the simulation eliminates the dead zone but does not significantly change the ejecta.
{\bf (a)} Good agreement between speed contours $|\v u|/U_0$ from discrete particle simulation with a frictional target (left, solid), the continuum frictional-fluid model with a no-slip target (left, dashed) and experiment (right). 
The grey curves outline the dead zone. 
{\bf (b)} Normalized scattering profiles from discrete-particle simulations for a frictional target (blue points) and a frictionless target (red points) show that the degree of collimation remains relatively unchanged regardless of the internal kinematic features.
}
\end{figure}

Fig.~\ref{fig:expdz} shows the azimuthally-averaged radial velocity along the smooth transparent target. 
Using particle tracking, we also measured the velocity fluctuations along the target base to obtain the normalized azimuthally-averaged granular temperature $\ave{T_{\rm eff}}/U_0^2$ (Fig.~\ref{fig:expdz} inset). The normalized temperature measures how much energy is contained within the velocity fluctuations relative to the kinetic energy of the incident jet. Its small numerical value shows that the kinetic energy originally possessed by the now immobile particles in the dead zone is largely dissipated. 
Fig.~\ref{fig:dem}{\bf (a)} shows speed contours in front of the roughened steel target, where $r$ and $z$ are the radial and axial coordinates. Defining a particle as being in the dead zone when $|\v u|^2/U_0^2 < 10^{-3}$, we see that this impact produces a dead zone that is broader and taller than the one obtained with a smooth target.  Specifically the dead-zone radius $R_{\rm DZ}$ is $0.76 R_{\rm Tar}$ while the height $H_{\rm DZ}$ is $0.4 R_{\rm Tar}$.

In the dilute jet limit, the ejection angle $\Psi_0$ of a single particle changes dramatically if the particle collides with a flat target instead of a conical dead-zone. While Cheng {\em et al}.~\cite{cheng} first noted that dense granular jet and water jet impact onto a flat target produces the same ejecta sheet angle $\Psi_0$, the experimental results presented here reveal that this similarity obtains in the presence of highly dissimilar internal structures.  Granular jet impact produces a dead zone while water jet impact creates an axisymmetric straining flow without a dead zone.  This makes the similar $\Psi_0$ values puzzling. 
 
To understand this, we recall the observation from Cheng {\em et al.} that, if the entire jet is modeled as a single degree-of-freedom system 
with an average velocity, in analogy with particle impact in the dilute regime, then momentum and energy conservation requires $\cos\Psi_0 = 1 - (A-B)(R_{\rm Tar}/R_{\rm Jet})^2$  when $R_{\rm Tar}<R_{\rm Jet}$~\cite{cheng,clanet}. The dimensionless constant $A$ is the reaction force exerted by the target, normalized by the incoming momentum flux $P_{\rm in}$ (defined as the total incoming jet momentum per unit time). The dimensionless constant $B$ relates $P_{\rm out}$, the momentum flux in the ejecta, via $P_{\rm out}/P_{\rm in} = 1 - B (R_{\rm Tar}/R_{\rm Jet})^2$. Because it quantifies the dissipation rate incurred during jet impact, $B$ is a normalized drag force.  In this single degree-of-freedom model, the fact that $\Psi_0$ is similar for water and granular jets means $A-B$ is similar and therefore the forces experienced during impact are alike. This is surprising, especially in light of our experimental finding that the internal flow is considerably different.


To explore the origin of the insensitivity of $\Psi_0$, we first constructed a discrete-particle simulation to track how some of the many degrees of freedom present in  dense granular impact evolve.  
In our scheme~\cite{nich_method}, the particles are modeled as hard spheres that experience dynamic friction.  Upon collision, the spheres lose a fraction of their kinetic energy. We consider two targets: (i) a frictional one where we decorate the target with a layer of stationary grains and (ii) a frictionless one where grains experience specular reflection upon collision with the target.

Frictional-target impact simulations reproduce the salient experimental results. As in the experiment, particles leave the target in a thin sheet while an approximately conical dead zone forms at the target. The normalized velocity contours from experiment using a roughened target and the frictional-target simulation agree quantitatively near the target in Fig.~\ref{fig:dem}{\bf (a)}.

\begin{figure}
\centering
\includegraphics[width=\columnwidth]{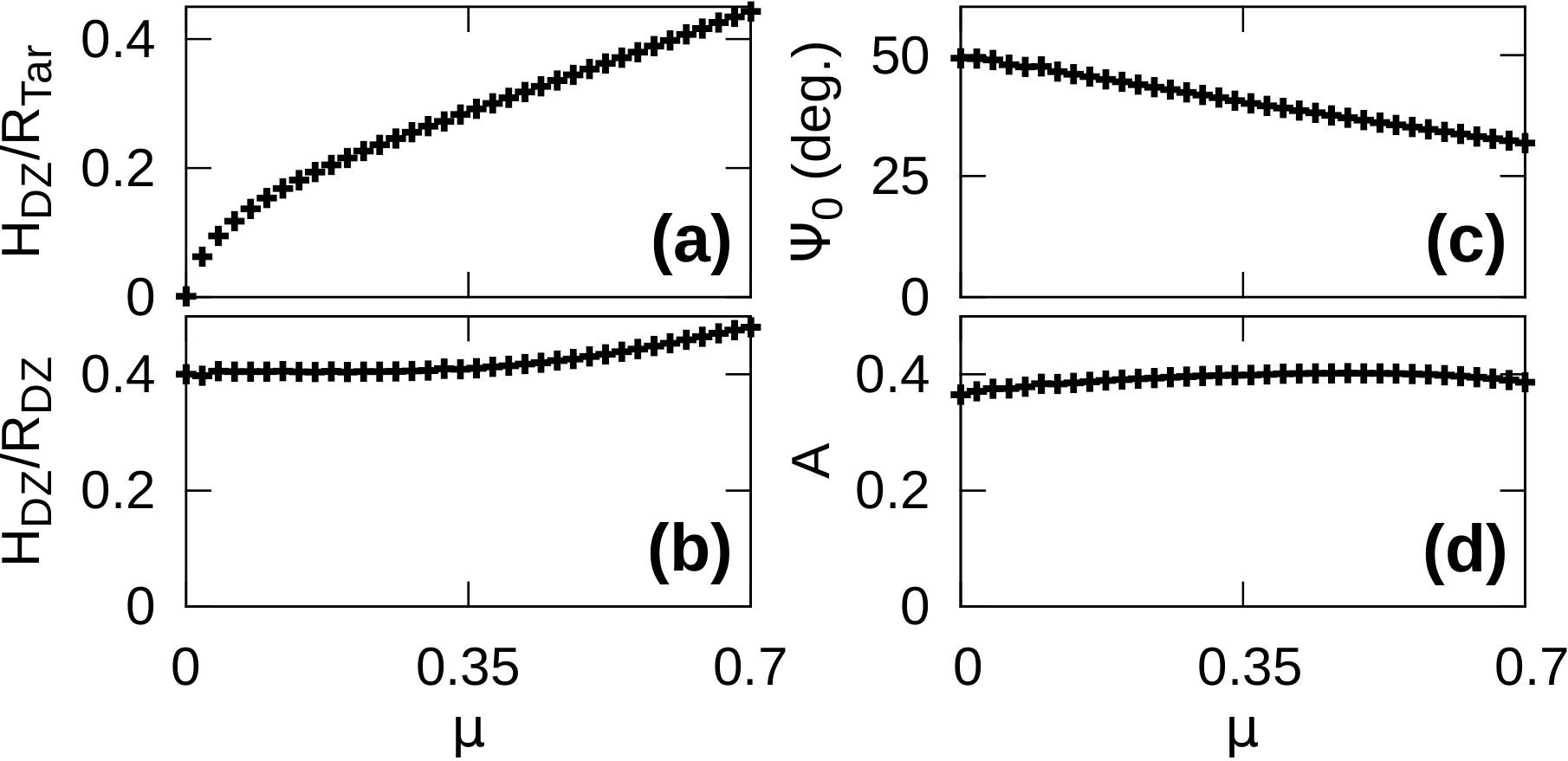}
\caption{\label{fig:cont} Solutions to 
the continuum granular-fluid model shows that changing the dead-zone size only affects the ejecta slightly.
{\bf (a)} As $\mu$ decreases, the normalized dead-zone height $H_{\rm DZ}/R_{\rm Tar}$ continuously shrinks and vanishes at $\mu=0$.
{\bf (b)} The dead-zone aspect ratio $H_{\rm DZ}/R_{\rm DZ}$ remains approximately constant for different dead-zone sizes.
{\bf (c)} The ejecta angle $\Psi_0$ changes only slightly even as the internal dead-zone shrinks from a significant region to $0$. 
{\bf (d)} The dimensionless reaction force exerted by the target on the jet $A$ barely changes over the entire range of $\mu$.
}
\end{figure}

Changing parameters, such as the coefficient of restitution or the friction between grains, produces only weak variations in the ejected sheet or the dead zone~\cite{nich_dissipation}. On the other hand, using frictionless targets instead of frictional ones produces a qualitative change: the dead zone is eliminated entirely. This dramatic change in the internal state of the jet produces only a slight change in the ejecta: $\Psi_0$ changes from $37^\circ$ for a frictional target to $45^\circ$ for a frictionless one.  The degree of collimation also remains similar (Fig.~\ref{fig:dem}{\bf (b)}).  These results show that the granular ejecta is remarkably insensitive to whether the impact zone is static or freely flowing but do indicate why granular jet impact behaves like water jet impact. 

To address this question, we analyze granular impact in the continuum limit.  Since both experiment and simulation show that the grains remain densely packed with a low effective temperature, we assume that the collective motion in the jet is incompressible and isothermal. In addition, we assume that the deviatoric stresses obey generalized Coulombic friction: they lie along the shear direction with a magnitude equal to the pressure multiplied by the dynamic-friction coefficient $\mu$. In contrast, Newtonian fluids, such as water, have deviatoric-stress components that are proportional to the strain rate but independent of pressure.

Incompressibility and momentum conservation then yield the following governing equations for the velocity field ${\v u}({\v x}, t)$ and the pressure field $p({\v x}, t)$
	\begin{align}
	&\boldsymbol\nabla\cdot\v u=0\\
	&\rho(\del_t + \v u\cdot\boldsymbol\nabla)\v u = \boldsymbol\nabla\cdot\boldsymbol\sigma, \quad \boldsymbol\sigma = -p\boldsymbol{\mathsf{I}} + \mu p\dot{\boldsymbol\gamma}/|\dot{\boldsymbol\gamma}|
	\end{align}
where $\boldsymbol\sigma$ is the stress tensor, $\boldsymbol{\mathsf{I}}$ is the identity matrix, $\dot{\boldsymbol\gamma} = \boldsymbol\nabla\v u + \boldsymbol\nabla\v u^{\rm T}$ is the rate of strain tensor, and $|\dot{\boldsymbol\gamma}| = \sqrt{(\dot{\boldsymbol\gamma}\boldsymbol:\dot{\boldsymbol\gamma})/2}$.  During steady impact, the {\it a priori} unknown jet surface satisfies the free stress condition $\boldsymbol\sigma\cdot\v n = \v 0$.  This boundary condition, together with a zero velocity condition at the target, completes the mathematical formulation of steady-state impact. 

Previous works~\cite{jop,muI,ecke,goldman_swim,goldman_swinney_impact} modeling dense granular flow as a frictional fluid have found the best agreement with experiments by allowing the dynamic friction $\mu$ to depend on a ratio of two timescales: a microscopic particle rearrange timescale determined by the local confining pressure, and a macroscopic timescale related to large-scale shear.  In general, wherever the pressure and velocity fields vary, this ratio varies as well.  Granular jet impact, however, is particularly simple because the local confining pressure is generated by impact alone and thus scales as $\rho U_0^2$.  As a result, the ratio of time-scales remains essentially uniform over the impact region.  We are therefore able to reproduce the experimental measurements by choosing $\mu$ to have a single value. 

We use the open-source, time-dependent free-surface solver Gerris~\cite{muI,gerris,gerris_online} to obtain steady-state solutions to the frictional fluid jet impact problem.  The solutions reproduce the pertinent features of the experiment and the discrete-particle simulations. Imposing no-slip boundary conditions at the target, so that the tangential speed of the fluid is zero there, creates a conical dead-zone within the jet. Imposing free-slip boundary conditions, corresponding to making the target frictionless, eliminates the dead zone. We chose $\mu=0.45$ to reproduce the experimental ejecta angle $\Psi_0$. This produces velocity fields that agree exceptionally well with experiment (Fig.~\ref{fig:dem}{\bf (a)}).

Fig.~\ref{fig:cont}{\bf (a)} shows that as $\mu$ decreases, so that the frictional-fluid model approaches perfect-fluid flow, the dead zone shrinks continuously and vanishes at $\mu=0$, while leaving $H_{\rm DZ}/R_{\rm DZ}$ approximately constant (Fig.~\ref{fig:cont}{\bf (b)}). In Fig.~\ref{fig:cont}{\bf (c)}, we observe that $\Psi_0$ only varies by $20^\circ$ even though the internal structure has changed dramatically. The variation is due almost entirely to the reduction in the normalized drag force $B$. The normalized reaction force $A$ is nearly constant in $\mu$, regardless of whether impact creates a sizable dead-zone or a vanishingly small one as shown in Fig.~\ref{fig:cont}{\bf (d)}. These results suggest that the insensitivity of the ejecta to internal structure is not due to dissipation or jet granularity; rather it owes its origin to the reaction force generated by perfect-fluid flow being insensitive to internal structure.

\begin{figure}
\centering
\includegraphics[width=\columnwidth]{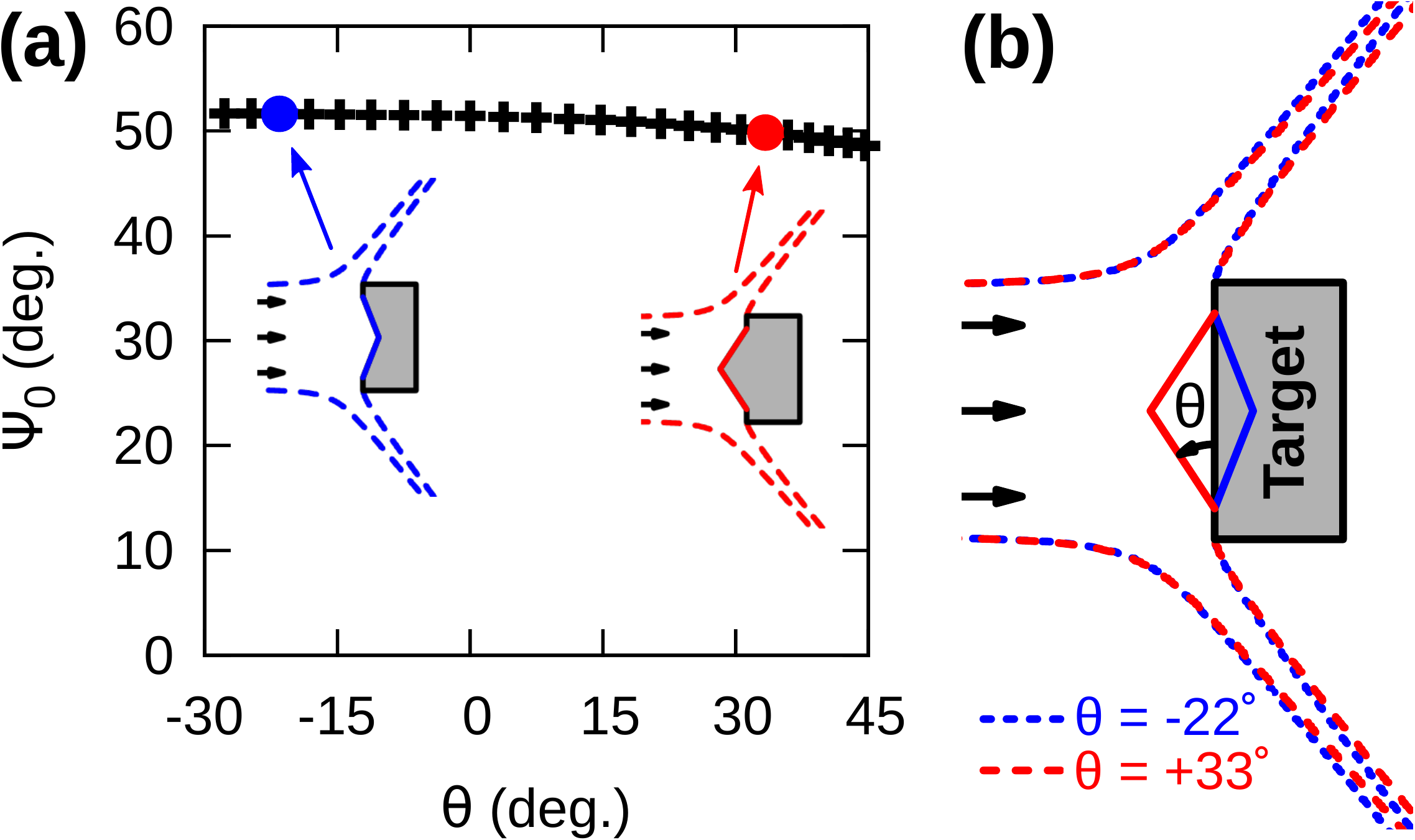}
\caption{\label{fig:euler}(color online) Perfect-fluid calculations demonstrate insensitivity of impact dynamics to the target structure. 
{\bf (a)} 
The ejecta angle $\Psi_0$ versus the inclination angle $\theta$ of a conical structure placed at the target center (Fig.~\ref{fig:euler} {\bf (b)}), where $\theta < 0$ (blue) is a depression into the target and $\theta > 0$ (red) is a conical dead-zone-like protrusion, demonstrate that even $\theta$ values considerably larger than the dead zone in our experiment show little signature in the ejecta.
{\bf (b)} Nearly identical jet surface cross-sections (dashed lines) for perfect-fluid impact against different targets (solid lines) nearly coincide.
}
\end{figure}

To confirm this hypothesis, we consider the impact of a perfect fluid onto a conical structure of inclination angle $\theta$. Positive $\theta$ is a protrusion, mimicking the dead-zone geometry, and negative $\theta$ is a depression. We match the cone-base radius to the experimental dead-zone radius, $0.76 R_{\rm Tar}$, and vary $\theta$ (Fig.~\ref{fig:euler}{\bf (a)}). The figure shows that ejecta sheets from three very different targets are nearly identical. Fig.~\ref{fig:euler}{\bf (b)} shows that changing $\theta$ between $-30^\circ$ and $+45^\circ$ produces a variation in the ejecta angle of merely $6$ percent. The experimental dead-zone corresponds to $\theta=+27^\circ$, within the range of $\theta$ presented.

Taken together, our results demonstrate that, absent direct measurements of internal states, a collimated ejecta pattern is just as plausibly produced by far-from-equilibrium collective motion as from a fully thermalized Newtonian flow. Here, collimated ejecta coexist with the creation of a cold dead-zone. Thus, the ejecta cannot be easily used to determine the internal state of the dense beam. This is relevant to the elliptical flow observed in the quark-gluon plasma at RHIC, which has been interpreted as evidence for fully thermalized Newtonian flow.

Our results are also relevant to accretion by dense granular impact. Even without cohesive forces between the frictional particles, jet impact accretes an interior dead-zone. This is relevant to the formation of planetesimals which have been difficult to model~\cite{blum_planetary,johansen_planetary,teiser_planetary}. 
Our finding supports the previously proposed view that the difference in porosity in colliding dust aggregates is more important than the precise strength of inter-particle cohesive forces~\cite{langkowski_teiser_blum,just_blum}.

In conclusion, we have investigated the relationship between the highly collimated, thin-sheet ejecta and the internal dynamics of a granular jet during impact. In contrast with recent theory and simulations by Sano and Hayakawa~\cite{sano1}, our experiment clearly shows that a large dead-zone forms on impact. Furthermore, our simulations show that thin ejecta sheets form generically when the effective temperature is low and the density is high, regardless of whether or not an interior dead-zone is present. Continuum modeling shows that the dynamics are well approximated as an incompressible frictional fluid. Changing the dissipation produces little variation in the ejecta and the target reaction force, even though the dead zone size varies greatly. This persists to the limiting case of a perfect fluid, which is also insensitive to the internal structure.  Thus the ejecta formed by granular impact is robust because they are dictated by incompressibility and inertia.

\begin{acknowledgments}
We thank Xiang Cheng and Heinrich M.~Jaeger for discussions. This research is supported by NSF MRSEC DMR-0820054 and NSF CBET-0967288. N.~G.~acknowledges support from a University of Chicago NSF MRSEC Kadanoff-Rice fellowship, and J.~E.~acknowledges support from a University of Chicago Millikan fellowship.
\end{acknowledgments}

\bibliography{deadzone}

\begin{thebibliography}{32}%
\makeatletter
\providecommand \@ifxundefined [1]{%
 \@ifx{#1\undefined}
}%
\providecommand \@ifnum [1]{%
 \ifnum #1\expandafter \@firstoftwo
 \else \expandafter \@secondoftwo
 \fi
}%
\providecommand \@ifx [1]{%
 \ifx #1\expandafter \@firstoftwo
 \else \expandafter \@secondoftwo
 \fi
}%
\providecommand \natexlab [1]{#1}%
\providecommand \enquote  [1]{``#1''}%
\providecommand \bibnamefont  [1]{#1}%
\providecommand \bibfnamefont [1]{#1}%
\providecommand \citenamefont [1]{#1}%
\providecommand \href@noop [0]{\@secondoftwo}%
\providecommand \href [0]{\begingroup \@sanitize@url \@href}%
\providecommand \@href[1]{\@@startlink{#1}\@@href}%
\providecommand \@@href[1]{\endgroup#1\@@endlink}%
\providecommand \@sanitize@url [0]{\catcode `\\12\catcode `\$12\catcode
  `\&12\catcode `\#12\catcode `\^12\catcode `\_12\catcode `\%12\relax}%
\providecommand \@@startlink[1]{}%
\providecommand \@@endlink[0]{}%
\providecommand \url  [0]{\begingroup\@sanitize@url \@url }%
\providecommand \@url [1]{\endgroup\@href {#1}{\urlprefix }}%
\providecommand \urlprefix  [0]{URL }%
\providecommand \Eprint [0]{\href }%
\providecommand \doibase [0]{http://dx.doi.org/}%
\providecommand \selectlanguage [0]{\@gobble}%
\providecommand \bibinfo  [0]{\@secondoftwo}%
\providecommand \bibfield  [0]{\@secondoftwo}%
\providecommand \translation [1]{[#1]}%
\providecommand \BibitemOpen [0]{}%
\providecommand \bibitemStop [0]{}%
\providecommand \bibitemNoStop [0]{.\EOS\space}%
\providecommand \EOS [0]{\spacefactor3000\relax}%
\providecommand \BibitemShut  [1]{\csname bibitem#1\endcsname}%
\let\auto@bib@innerbib\@empty
\bibitem [{\citenamefont {Chandler}\ \emph {et~al.}(1983)\citenamefont
  {Chandler}, \citenamefont {Weeks},\ and\ \citenamefont
  {Andersen}}]{van_der_waals}%
  \BibitemOpen
  \bibfield  {author} {\bibinfo {author} {\bibfnamefont {D.}~\bibnamefont
  {Chandler}}, \bibinfo {author} {\bibfnamefont {J.~D.}\ \bibnamefont {Weeks}},
  \ and\ \bibinfo {author} {\bibfnamefont {H.~C.}\ \bibnamefont {Andersen}},\
  }\href@noop {} {\bibfield  {journal} {\bibinfo  {journal} {Science}\ }\textbf
  {\bibinfo {volume} {220}},\ \bibinfo {pages} {787} (\bibinfo {year}
  {1983})}\BibitemShut {NoStop}%
\bibitem [{\citenamefont {Longuet-Higgins}\ and\ \citenamefont
  {Widom}(1964)}]{rigid_sphere}%
  \BibitemOpen
  \bibfield  {author} {\bibinfo {author} {\bibfnamefont {H.~C.}\ \bibnamefont
  {Longuet-Higgins}}\ and\ \bibinfo {author} {\bibfnamefont {B.}~\bibnamefont
  {Widom}},\ }\href@noop {} {\bibfield  {journal} {\bibinfo  {journal} {Mol.
  Phys.}\ }\textbf {\bibinfo {volume} {8}},\ \bibinfo {pages} {549} (\bibinfo
  {year} {1964})}\BibitemShut {NoStop}%
\bibitem [{\citenamefont {Orpe}\ and\ \citenamefont
  {Kudrolli}(2007)}]{kudrolli}%
  \BibitemOpen
  \bibfield  {author} {\bibinfo {author} {\bibfnamefont {A.}~\bibnamefont
  {Orpe}}\ and\ \bibinfo {author} {\bibfnamefont {A.}~\bibnamefont
  {Kudrolli}},\ }\href@noop {} {\bibfield  {journal} {\bibinfo  {journal}
  {Phys. Rev. Lett.}\ }\textbf {\bibinfo {volume} {98}},\ \bibinfo {pages}
  {238001} (\bibinfo {year} {2007})}\BibitemShut {NoStop}%
\bibitem [{\citenamefont {Cheng}\ \emph {et~al.}(2007)\citenamefont {Cheng},
  \citenamefont {Varas}, \citenamefont {Citron}, \citenamefont {Jaeger},\ and\
  \citenamefont {Nagel}}]{cheng}%
  \BibitemOpen
  \bibfield  {author} {\bibinfo {author} {\bibfnamefont {X.}~\bibnamefont
  {Cheng}}, \bibinfo {author} {\bibfnamefont {G.}~\bibnamefont {Varas}},
  \bibinfo {author} {\bibfnamefont {D.}~\bibnamefont {Citron}}, \bibinfo
  {author} {\bibfnamefont {H.~M.}\ \bibnamefont {Jaeger}}, \ and\ \bibinfo
  {author} {\bibfnamefont {S.~R.}\ \bibnamefont {Nagel}},\ }\href@noop {}
  {\bibfield  {journal} {\bibinfo  {journal} {Phys. Rev. Lett.}\ }\textbf
  {\bibinfo {volume} {99}},\ \bibinfo {pages} {188001} (\bibinfo {year}
  {2007})}\BibitemShut {NoStop}%
\bibitem [{\citenamefont {Clanet}(2001)}]{clanet}%
  \BibitemOpen
  \bibfield  {author} {\bibinfo {author} {\bibfnamefont {C.}~\bibnamefont
  {Clanet}},\ }\href@noop {} {\bibfield  {journal} {\bibinfo  {journal} {J.
  Fluid Mech.}\ }\textbf {\bibinfo {volume} {430}},\ \bibinfo {pages} {111}
  (\bibinfo {year} {2001})}\BibitemShut {NoStop}%
\bibitem [{\citenamefont {Braun-Munzinger}\ and\ \citenamefont
  {Satchel}(2007)}]{rhic_nature}%
  \BibitemOpen
  \bibfield  {author} {\bibinfo {author} {\bibfnamefont {P.}~\bibnamefont
  {Braun-Munzinger}}\ and\ \bibinfo {author} {\bibfnamefont {J.}~\bibnamefont
  {Satchel}},\ }\href@noop {} {\bibfield  {journal} {\bibinfo  {journal}
  {Nature}\ }\textbf {\bibinfo {volume} {448}},\ \bibinfo {pages} {302}
  (\bibinfo {year} {2007})}\BibitemShut {NoStop}%
\bibitem [{\citenamefont {Romatschke}\ and\ \citenamefont
  {Romatschke}(2009)}]{rhic_euler}%
  \BibitemOpen
  \bibfield  {author} {\bibinfo {author} {\bibfnamefont {P.}~\bibnamefont
  {Romatschke}}\ and\ \bibinfo {author} {\bibfnamefont {U.}~\bibnamefont
  {Romatschke}},\ }\href@noop {} {\bibfield  {journal} {\bibinfo  {journal}
  {Phys. Rev. Lett.}\ }\textbf {\bibinfo {volume} {99}},\ \bibinfo {pages}
  {172301} (\bibinfo {year} {2009})}\BibitemShut {NoStop}%
\bibitem [{\citenamefont {Song}\ and\ \citenamefont
  {Heinz}(2009)}]{song_viscosity}%
  \BibitemOpen
  \bibfield  {author} {\bibinfo {author} {\bibfnamefont {H.}~\bibnamefont
  {Song}}\ and\ \bibinfo {author} {\bibfnamefont {U.}~\bibnamefont {Heinz}},\
  }\href@noop {} {\bibfield  {journal} {\bibinfo  {journal} {J. Phys. G}\
  }\textbf {\bibinfo {volume} {36}},\ \bibinfo {pages} {064033} (\bibinfo
  {year} {2009})}\BibitemShut {NoStop}%
\bibitem [{\citenamefont {Jacak}\ and\ \citenamefont
  {Steinberg}(2010)}]{jacak_rhic}%
  \BibitemOpen
  \bibfield  {author} {\bibinfo {author} {\bibfnamefont {B.}~\bibnamefont
  {Jacak}}\ and\ \bibinfo {author} {\bibfnamefont {P.}~\bibnamefont
  {Steinberg}},\ }\href@noop {} {\bibfield  {journal} {\bibinfo  {journal}
  {Phys. Today}\ }\textbf {\bibinfo {volume} {63}} (\bibinfo {year}
  {2010})}\BibitemShut {NoStop}%
\bibitem [{\citenamefont {Jaeger}\ and\ \citenamefont
  {Nagel}(1992)}]{jaeger_sandpile}%
  \BibitemOpen
  \bibfield  {author} {\bibinfo {author} {\bibfnamefont {H.~M.}\ \bibnamefont
  {Jaeger}}\ and\ \bibinfo {author} {\bibfnamefont {S.~R.}\ \bibnamefont
  {Nagel}},\ }\href@noop {} {\bibfield  {journal} {\bibinfo  {journal}
  {Science}\ }\textbf {\bibinfo {volume} {255}},\ \bibinfo {pages} {1523}
  (\bibinfo {year} {1992})}\BibitemShut {NoStop}%
\bibitem [{\citenamefont {Lube}\ \emph {et~al.}(2004)\citenamefont {Lube},
  \citenamefont {Huppert}, \citenamefont {Sparks},\ and\ \citenamefont
  {Hallworth}}]{lube_huppert_sandpile}%
  \BibitemOpen
  \bibfield  {author} {\bibinfo {author} {\bibfnamefont {G.}~\bibnamefont
  {Lube}}, \bibinfo {author} {\bibfnamefont {H.~E.}\ \bibnamefont {Huppert}},
  \bibinfo {author} {\bibfnamefont {R.~S.~J.}\ \bibnamefont {Sparks}}, \ and\
  \bibinfo {author} {\bibfnamefont {M.~A.}\ \bibnamefont {Hallworth}},\
  }\href@noop {} {\bibfield  {journal} {\bibinfo  {journal} {J. Fluid Mech.}\
  }\textbf {\bibinfo {volume} {508}},\ \bibinfo {pages} {175} (\bibinfo {year}
  {2004})}\BibitemShut {NoStop}%
\bibitem [{\citenamefont {Amarouchene}\ \emph {et~al.}(2001)\citenamefont
  {Amarouchene}, \citenamefont {Boudet},\ and\ \citenamefont
  {Kellay}}]{sand_dune}%
  \BibitemOpen
  \bibfield  {author} {\bibinfo {author} {\bibfnamefont {Y.}~\bibnamefont
  {Amarouchene}}, \bibinfo {author} {\bibfnamefont {F.}~\bibnamefont {Boudet}},
  \ and\ \bibinfo {author} {\bibfnamefont {H.}~\bibnamefont {Kellay}},\
  }\href@noop {} {\bibfield  {journal} {\bibinfo  {journal} {Phys. Rev. Lett.}\
  }\textbf {\bibinfo {volume} {86}},\ \bibinfo {pages} {4286} (\bibinfo {year}
  {2001})}\BibitemShut {NoStop}%
\bibitem [{\citenamefont {Gray}\ \emph {et~al.}(2003)\citenamefont {Gray},
  \citenamefont {Tai},\ and\ \citenamefont {Noelle}}]{gray_shock}%
  \BibitemOpen
  \bibfield  {author} {\bibinfo {author} {\bibfnamefont {J.~M.~N.~T.}\
  \bibnamefont {Gray}}, \bibinfo {author} {\bibfnamefont {Y.-C.}\ \bibnamefont
  {Tai}}, \ and\ \bibinfo {author} {\bibfnamefont {S.}~\bibnamefont {Noelle}},\
  }\href@noop {} {\bibfield  {journal} {\bibinfo  {journal} {J. Fluid Mech.}\
  }\textbf {\bibinfo {volume} {491}},\ \bibinfo {pages} {161} (\bibinfo {year}
  {2003})}\BibitemShut {NoStop}%
\bibitem [{\citenamefont {Royer}\ \emph {et~al.}(2009)\citenamefont {Royer},
  \citenamefont {Evans}, \citenamefont {Oyarte}, \citenamefont {Guo},
  \citenamefont {Kapit}, \citenamefont {Mobius}, \citenamefont {Waitukaitis},\
  and\ \citenamefont {Jaeger}}]{granular_stream}%
  \BibitemOpen
  \bibfield  {author} {\bibinfo {author} {\bibfnamefont {J.~R.}\ \bibnamefont
  {Royer}}, \bibinfo {author} {\bibfnamefont {D.~J.}\ \bibnamefont {Evans}},
  \bibinfo {author} {\bibfnamefont {L.}~\bibnamefont {Oyarte}}, \bibinfo
  {author} {\bibfnamefont {Q.}~\bibnamefont {Guo}}, \bibinfo {author}
  {\bibfnamefont {E.}~\bibnamefont {Kapit}}, \bibinfo {author} {\bibfnamefont
  {M.~E.}\ \bibnamefont {Mobius}}, \bibinfo {author} {\bibfnamefont {S.~R.}\
  \bibnamefont {Waitukaitis}}, \ and\ \bibinfo {author} {\bibfnamefont {H.~M.}\
  \bibnamefont {Jaeger}},\ }\href@noop {} {\bibfield  {journal} {\bibinfo
  {journal} {Nature}\ }\textbf {\bibinfo {volume} {459}},\ \bibinfo {pages}
  {1110} (\bibinfo {year} {2009})}\BibitemShut {NoStop}%
\bibitem [{\citenamefont {Rericha}\ \emph {et~al.}(2001)\citenamefont
  {Rericha}, \citenamefont {Bizon}, \citenamefont {Shattuck},\ and\
  \citenamefont {Swinney}}]{swinney_shock}%
  \BibitemOpen
  \bibfield  {author} {\bibinfo {author} {\bibfnamefont {E.~C.}\ \bibnamefont
  {Rericha}}, \bibinfo {author} {\bibfnamefont {C.}~\bibnamefont {Bizon}},
  \bibinfo {author} {\bibfnamefont {M.~D.}\ \bibnamefont {Shattuck}}, \ and\
  \bibinfo {author} {\bibfnamefont {H.~L.}\ \bibnamefont {Swinney}},\
  }\href@noop {} {\bibfield  {journal} {\bibinfo  {journal} {Phys. Rev. Lett.}\
  }\textbf {\bibinfo {volume} {88}},\ \bibinfo {pages} {014302} (\bibinfo
  {year} {2001})}\BibitemShut {NoStop}%
\bibitem [{\citenamefont {Blum}\ and\ \citenamefont
  {Wurm}(2008)}]{blum_planetary}%
  \BibitemOpen
  \bibfield  {author} {\bibinfo {author} {\bibfnamefont {J.}~\bibnamefont
  {Blum}}\ and\ \bibinfo {author} {\bibfnamefont {G.}~\bibnamefont {Wurm}},\
  }\href@noop {} {\bibfield  {journal} {\bibinfo  {journal} {Ann. Rev. Astron.
  Astrophys.}\ }\textbf {\bibinfo {volume} {46}},\ \bibinfo {pages} {21}
  (\bibinfo {year} {2008})}\BibitemShut {NoStop}%
\bibitem [{\citenamefont {Johansen}\ \emph {et~al.}(2007)\citenamefont
  {Johansen}, \citenamefont {Oishi}, \citenamefont {Low}, \citenamefont
  {Klahr}, \citenamefont {Henning},\ and\ \citenamefont
  {Youdin}}]{johansen_planetary}%
  \BibitemOpen
  \bibfield  {author} {\bibinfo {author} {\bibfnamefont {A.}~\bibnamefont
  {Johansen}}, \bibinfo {author} {\bibfnamefont {J.~S.}\ \bibnamefont {Oishi}},
  \bibinfo {author} {\bibfnamefont {M.-M.~M.}\ \bibnamefont {Low}}, \bibinfo
  {author} {\bibfnamefont {H.}~\bibnamefont {Klahr}}, \bibinfo {author}
  {\bibfnamefont {T.}~\bibnamefont {Henning}}, \ and\ \bibinfo {author}
  {\bibfnamefont {A.}~\bibnamefont {Youdin}},\ }\href@noop {} {\bibfield
  {journal} {\bibinfo  {journal} {Nature}\ }\textbf {\bibinfo {volume} {448}},\
  \bibinfo {pages} {1022} (\bibinfo {year} {2007})}\BibitemShut {NoStop}%
\bibitem [{\citenamefont {Teiser}\ and\ \citenamefont
  {Wurm}(2009)}]{teiser_planetary}%
  \BibitemOpen
  \bibfield  {author} {\bibinfo {author} {\bibfnamefont {J.}~\bibnamefont
  {Teiser}}\ and\ \bibinfo {author} {\bibfnamefont {G.}~\bibnamefont {Wurm}},\
  }\href@noop {} {\bibfield  {journal} {\bibinfo  {journal} {Mon. Not. R.
  Astron. Soc.}\ }\textbf {\bibinfo {volume} {393}},\ \bibinfo {pages} {1584}
  (\bibinfo {year} {2009})}\BibitemShut {NoStop}%
\bibitem [{\citenamefont {Blum}(2010)}]{just_blum}%
  \BibitemOpen
  \bibfield  {author} {\bibinfo {author} {\bibfnamefont {J.}~\bibnamefont
  {Blum}},\ }\href@noop {} {\bibfield  {journal} {\bibinfo  {journal} {Res.
  Astron. Astrophys.}\ }\textbf {\bibinfo {volume} {10}},\ \bibinfo {pages}
  {1199} (\bibinfo {year} {2010})}\BibitemShut {NoStop}%
\bibitem [{\citenamefont {Langkowski}\ \emph {et~al.}(2008)\citenamefont
  {Langkowski}, \citenamefont {Teiser},\ and\ \citenamefont
  {Blum}}]{langkowski_teiser_blum}%
  \BibitemOpen
  \bibfield  {author} {\bibinfo {author} {\bibfnamefont {D.}~\bibnamefont
  {Langkowski}}, \bibinfo {author} {\bibfnamefont {J.}~\bibnamefont {Teiser}},
  \ and\ \bibinfo {author} {\bibfnamefont {J.}~\bibnamefont {Blum}},\
  }\href@noop {} {\bibfield  {journal} {\bibinfo  {journal} {Astrophysic. J.}\
  }\textbf {\bibinfo {volume} {675}},\ \bibinfo {pages} {764} (\bibinfo {year}
  {2008})}\BibitemShut {NoStop}%
\bibitem [{\citenamefont {Seavey}(1986)}]{sandblast1}%
  \BibitemOpen
  \bibfield  {author} {\bibinfo {author} {\bibfnamefont {M.}~\bibnamefont
  {Seavey}},\ }\href@noop {} {\bibfield  {journal} {\bibinfo  {journal} {J.
  Prot. Coatings and Linings}\ }\textbf {\bibinfo {volume} {2}},\ \bibinfo
  {pages} {26} (\bibinfo {year} {1986})}\BibitemShut {NoStop}%
\bibitem [{\citenamefont {Achtsnicka}\ \emph {et~al.}(2005)\citenamefont
  {Achtsnicka}, \citenamefont {Geelhoedb}, \citenamefont {Hoogstratea},\ and\
  \citenamefont {Karpuschewski}}]{sandblast2}%
  \BibitemOpen
  \bibfield  {author} {\bibinfo {author} {\bibfnamefont {M.}~\bibnamefont
  {Achtsnicka}}, \bibinfo {author} {\bibfnamefont {P.~F.}\ \bibnamefont
  {Geelhoedb}}, \bibinfo {author} {\bibfnamefont {A.~M.}\ \bibnamefont
  {Hoogstratea}}, \ and\ \bibinfo {author} {\bibfnamefont {B.}~\bibnamefont
  {Karpuschewski}},\ }\href@noop {} {\bibfield  {journal} {\bibinfo  {journal}
  {Wear}\ }\textbf {\bibinfo {volume} {259}},\ \bibinfo {pages} {84} (\bibinfo
  {year} {2005})}\BibitemShut {NoStop}%
\bibitem [{\citenamefont {Guttenberg}(2011)}]{nich_method}%
  \BibitemOpen
  \bibfield  {author} {\bibinfo {author} {\bibfnamefont {N.}~\bibnamefont
  {Guttenberg}},\ }\href@noop {} {\bibfield  {journal} {\bibinfo  {journal}
  {Phys. Rev. E}\ }\textbf {\bibinfo {volume} {83}},\ \bibinfo {pages} {051306}
  (\bibinfo {year} {2011})}\BibitemShut {NoStop}%
\bibitem [{\citenamefont {Guttenberg}(2012)}]{nich_dissipation}%
  \BibitemOpen
  \bibfield  {author} {\bibinfo {author} {\bibfnamefont {N.}~\bibnamefont
  {Guttenberg}},\ }\href@noop {} {\bibfield  {journal} {\bibinfo  {journal}
  {Phys. Rev. E}\ }\textbf {\bibinfo {volume} {85}},\ \bibinfo {pages} {051303}
  (\bibinfo {year} {2012})}\BibitemShut {NoStop}%
\bibitem [{\citenamefont {Jop}\ \emph {et~al.}(2006)\citenamefont {Jop},
  \citenamefont {Forterre},\ and\ \citenamefont {Pouliquen}}]{jop}%
  \BibitemOpen
  \bibfield  {author} {\bibinfo {author} {\bibfnamefont {P.}~\bibnamefont
  {Jop}}, \bibinfo {author} {\bibfnamefont {Y.}~\bibnamefont {Forterre}}, \
  and\ \bibinfo {author} {\bibfnamefont {O.}~\bibnamefont {Pouliquen}},\
  }\href@noop {} {\bibfield  {journal} {\bibinfo  {journal} {Nature}\ }\textbf
  {\bibinfo {volume} {441}},\ \bibinfo {pages} {727} (\bibinfo {year}
  {2006})}\BibitemShut {NoStop}%
\bibitem [{\citenamefont {Lagre'e}\ \emph {et~al.}(2011)\citenamefont
  {Lagre'e}, \citenamefont {Staron},\ and\ \citenamefont {Popinet}}]{muI}%
  \BibitemOpen
  \bibfield  {author} {\bibinfo {author} {\bibfnamefont {P.-Y.}\ \bibnamefont
  {Lagre'e}}, \bibinfo {author} {\bibfnamefont {L.}~\bibnamefont {Staron}}, \
  and\ \bibinfo {author} {\bibfnamefont {S.}~\bibnamefont {Popinet}},\
  }\href@noop {} {\bibfield  {journal} {\bibinfo  {journal} {J. Fluid Mech.}\
  }\textbf {\bibinfo {volume} {686}},\ \bibinfo {pages} {378} (\bibinfo {year}
  {2011})}\BibitemShut {NoStop}%
\bibitem [{\citenamefont {Borzsonyi}\ and\ \citenamefont {Ecke}(2007)}]{ecke}%
  \BibitemOpen
  \bibfield  {author} {\bibinfo {author} {\bibfnamefont {T.}~\bibnamefont
  {Borzsonyi}}\ and\ \bibinfo {author} {\bibfnamefont {R.~E.}\ \bibnamefont
  {Ecke}},\ }\href@noop {} {\bibfield  {journal} {\bibinfo  {journal} {Phys.
  Rev. E}\ }\textbf {\bibinfo {volume} {76}},\ \bibinfo {pages} {031301}
  (\bibinfo {year} {2007})}\BibitemShut {NoStop}%
\bibitem [{\citenamefont {Maladen}\ \emph {et~al.}(2009)\citenamefont
  {Maladen}, \citenamefont {Ding}, \citenamefont {Li},\ and\ \citenamefont
  {Goldman}}]{goldman_swim}%
  \BibitemOpen
  \bibfield  {author} {\bibinfo {author} {\bibfnamefont {R.~D.}\ \bibnamefont
  {Maladen}}, \bibinfo {author} {\bibfnamefont {Y.}~\bibnamefont {Ding}},
  \bibinfo {author} {\bibfnamefont {C.}~\bibnamefont {Li}}, \ and\ \bibinfo
  {author} {\bibfnamefont {D.~I.}\ \bibnamefont {Goldman}},\ }\href@noop {}
  {\bibfield  {journal} {\bibinfo  {journal} {Science}\ }\textbf {\bibinfo
  {volume} {325}},\ \bibinfo {pages} {314} (\bibinfo {year}
  {2009})}\BibitemShut {NoStop}%
\bibitem [{\citenamefont {Ciamarra}\ \emph {et~al.}(2004)\citenamefont
  {Ciamarra}, \citenamefont {Lara}, \citenamefont {Lee}, \citenamefont
  {Goldman},\ and\ \citenamefont {Swinney}}]{goldman_swinney_impact}%
  \BibitemOpen
  \bibfield  {author} {\bibinfo {author} {\bibfnamefont {M.~P.}\ \bibnamefont
  {Ciamarra}}, \bibinfo {author} {\bibfnamefont {A.~H.}\ \bibnamefont {Lara}},
  \bibinfo {author} {\bibfnamefont {A.~T.}\ \bibnamefont {Lee}}, \bibinfo
  {author} {\bibfnamefont {D.~I.}\ \bibnamefont {Goldman}}, \ and\ \bibinfo
  {author} {\bibfnamefont {H.~L.}\ \bibnamefont {Swinney}},\ }\href@noop {}
  {\bibfield  {journal} {\bibinfo  {journal} {Phys. Rev. Lett.}\ }\textbf
  {\bibinfo {volume} {92}},\ \bibinfo {pages} {194301} (\bibinfo {year}
  {2004})}\BibitemShut {NoStop}%
\bibitem [{\citenamefont {Popinet}(2009)}]{gerris}%
  \BibitemOpen
  \bibfield  {author} {\bibinfo {author} {\bibfnamefont {S.}~\bibnamefont
  {Popinet}},\ }\href@noop {} {\bibfield  {journal} {\bibinfo  {journal} {J.
  Comp. Phys.}\ }\textbf {\bibinfo {volume} {228}},\ \bibinfo {pages} {5838}
  (\bibinfo {year} {2009})}\BibitemShut {NoStop}%
\bibitem [{ger()}]{gerris_online}%
  \BibitemOpen
  \href@noop {} {}\bibinfo {howpublished} {Available at
  \url{gfs.sf.net}}\BibitemShut {NoStop}%
\bibitem [{\citenamefont {Sano}\ and\ \citenamefont {Hayakawa}(2012)}]{sano1}%
  \BibitemOpen
  \bibfield  {author} {\bibinfo {author} {\bibfnamefont {T.~G.}\ \bibnamefont
  {Sano}}\ and\ \bibinfo {author} {\bibfnamefont {H.}~\bibnamefont
  {Hayakawa}},\ }\href@noop {} {\bibfield  {journal} {\bibinfo  {journal}
  {Phys. Rev. E}\ }\textbf {\bibinfo {volume} {86}},\ \bibinfo {pages} {041308}
  (\bibinfo {year} {2012})}\BibitemShut {NoStop}%
\end{thebibliography}%

\end{document}